\documentclass{iaus}
\usepackage{graphicx}

\title[CATS] 
{CATS: CfAO  Treasury Survey of distant galaxies, supernovae, and  AGN's}

\author[Koo et al. ]   
{David C. Koo$^{1,4}$, %
 Jason Melbourne$^1$, 
 Claire Max$^{1,4}$,
 Anne Metevier$^{1,4}$,
 Mark Ammons$^1$,
 James E. Larkin$^2$,
 Matthew Barczys$^2$,
 Shelley A. Wright $^2$,
 \and Eric Steinbring $^3$}

\affiliation{
$^1$Dept. of Astronomy and Astrophysics,  University of California,
Santa Cruz, CA 95064, USA  \break
email: ammons@ucolick.org, anne@ucolick.org, jmel@ucolick.org, koo@ucolick.org, max@ucolick.org\\[\affilskip]
$^2$ Dept. of Physics and Astronomy, University of California, Los Angeles, CA 90095, USA \break email: barczysm@astro.ucla.edu, larkin@astro.ucla.edu, saw@astro.ucla.edu \\[\affilskip]
$^3$ Herzberg Institute of Astrophysics, NRC Canada, Victoria, BC V9E 2E7, Canada  \break
 email: Eric.Steinbring@nrc-cnrc.gc.ca \\[\affilskip]
 $^4$UCO/Lick Observatory, University of California,
Santa Cruz, CA 95064, USA 
}
 
\pubyear{2006}
\volume{235}  
\pagerange{119--126}
\date{?? and in revised form ??}
\jname{Galaxy Evolution across the Hubble Time}
\editors{F.Combes \& J. Palous, eds.}
\begin{document}

\maketitle

\begin{abstract}
The NSF Science and Technology Center for Adaptive Optics (CfAO) is supporting a major scientific legacy project  called the CfAO Treasury Survey (CATS). CATS is obtaining near-infrared AO data in  deep HST survey fields, such as GEMS, GOODS-N, \& EGS.  Besides summarizing  the main objectives of CATS, we highlight some recent imaging  work  on the study of distant field galaxies, AGNs, and a redshift $z = 1.32$ supernova. 
CATS plans the first data release to the community in early 2007 
(check http://www.astro.ucla.edu/$\sim$irlab/cats/index.shtml for more details on CATS and latest updates).

\keywords{Instrument: adaptive optics; surveys; galaxies: active; galaxies: evolution; supernovae}
\end{abstract}
\firstsection 
\section{Why CATS?}
CATS is a long-term survey aimed for an enhanced near-IR study of the structural, chemical, star formation, and kinematic evolution of distant field galaxy subcomponents on sub-kpc scales. This is possible by using adaptive optics (AO) on an 8-10m class telescope  that has  diffraction-limits in the near-IR of 0.05 arcsec, which is  3x-4x better than is possible with HSTÕs 2.4m mirror at a similar wavelength.  This resolution  corresponds to the sizes of  bulges, disks, bars, spiral arms, and merger/interaction or lensing signatures at redshifts  $z > 0.5$ and improves the study of {\it unresolved} star forming knots, supernovae, and AGN's.  It also matches the optical diffraction-limit  of HST, so the CATS  strategy is to work  in well-studied HST survey  fields with ACS images taken in 2 or more filters, including GOODS, GEMS, \& EGS.  These regions are highly leveraged  by deep Chandra, XMM-Newton, GALEX,  and Spitzer data from space   and optical, near-IR, submm, and radio data from the ground.  

Keck's AO is presently limited to the  near-IR, which complements  the optical by providing superior  penetration of  dust-obscured regions and having higher sensitivity  to old stars. For high redshift objects, the near-IR measures   light that was emitted as restframe optical, including the important $H\alpha$ and [NII] lines for spectroscopy.  Keck II now routinely provides laser guide star (LGS)  AO (\cite[Wizinowich \etal\ 2006]{Wizinowich06}), which has opened new doors for many areas of astronomy  (see \cite[Liu 2006]{Liu06} for an overview of the advantages, disadvantages, and potential of AO).  With LGS AO supporting  tip-tilt stars as faint as  18th mag within $\sim60$ arcsec (\cite[van Dam \etal\ 2006]{vandam06}) and good characterization of the off-axis point spread function (\cite[Steinbring \etal\ 2005]{Steinbring05}), CATS now  has, for the first time, access to 20\% or more  of  the  area of the special  HST  galaxy survey fields, and potentially over 1500 galaxies to be studied with AO.   Prior to LGS, only about   1\%  was reachable on Keck since much brighter natural guide stars of 12th mag were needed (\cite[Wizinowich \etal\ 2000]{Wizinowich00}).  Thus much of the early distant galaxy research with Keck AO by \cite{Larkin00} and \cite{Glassman02} were undertaken without having the benefit of adding HST data (c.f., \cite[Steinbring \etal\ 2004]{Steinbring04}).

\section{Highlights}\label{highlights}
To date, CATS has obtained  AO data with Keck II for about 10 pointings or 5 square arcmin with  the NIRC2 camera in GOODS, GEMS, and EGS.   One Chandra X-ray galaxy  in GOODS-S was found to have two red nuclei  for which the 4 filter HST ACS photometry in the optical indicated either old stars or dusty, younger stars. The AO K-band  photometry provided the critical discrimination and suggested that both were old. Thus we were witnessing a "dry merger" (\cite[Melbourne \etal\ 2005]{Melbourne05}). 
In another study, M. Barczys (PhD 2006) exploited the higher sensitivity to stellar mass of near-IR photometry to discover that many of the apparent major mergers with roughly equal luminosities seen in the HST optical images are in fact minor mergers with unequal stellar masses.  Finally, CATS has recently achieved an important  milestone by measuring the very faint ($\sim$ 24 mag Vega) H band flux (and thus the  restframe R luminosity)  of  a high redshift ($z = 1.32$) supernova that was lying atop its much brighter host galaxy (\cite[Melbourne \etal\ 2007]{Melbourne07}).  

CATS plans to acquire more LGS AO images, mainly  to study the host galaxies of AGNs and  the colors of bulges and disks. With the advent of OSIRIS (\cite[Larkin \etal\ 2006]{Larkin06}), an AO compatible near-IR spectrograph with an integral field unit, CATS is poised to gather high spatial resolution 2-D  spectroscopic data.   A major goal of CATS is to provide the community with good quality AO data with which to explore its scientific potential. To this end, CATS intends to release some  Keck AO data by early 2007.  

\begin{acknowledgments}
We thank
the Keck Observatory staff for years of  help and for making LGS AO  a reality. We thank S. Perlmutter and  the Supernovae Cosmology Project  team  for providing  CATS with the supernova.  This work was supported in part by the NSF Science and Technology CfAO, managed by UC Santa Cruz under cooperative agreement No. AST-9876783. We close with thanks  to the Hawaiian people for use of their sacred mountain.
\end{acknowledgments}

\end{document}